\documentstyle[aps]{revtex}
\begin{document}
\title{ Einstein-Podolsky-Rosen Paradox and Antiparticle}
\author{Guang-jiong Ni\footnote{E-mail: gjni@fudan.ac.cn}} 
\address{Department of Physics, Fudan University, Shanghai, 200433,
  P. R. China}
\author{Hong Guan\footnote{E-mail: stdp@zsulink.zsu.edu.cn}}
\address{Department of Physics, Zhongshan University, Guangzhou,
  510275, P. R. China}
\maketitle
\begin{abstract}
The original version of Einstein-Podolsky-Rosen (EPR) paradox is discussed
to show the completeness of Quantum Mechanics (QM). The unique solution
leads to the wave function of antiparticle unambiguously, which implies the
essential conformity between QM and Special Relativity (SR).

PACS: 03.65.Bz
\end{abstract}

Einstein, Podolsky and Rosen (EPR) in their famous paper in 1935
\cite{r1} insisted on querying that ``Is the description of physical reality
in quantum mechanics (QM) complete?'' What EPR were discussing is a system
composing of two spinless particles. It was a very strange example\cite{r2},
see below. So quite naturally, beginning from Bohm \cite{r3}, physicists
have been turning to other experiments involving particle spin or photons.
The investigations on these EPR experiments together with the analysis of
Bell's inequality \cite{r4} have been revealing that the prediction of QM is
correct while the existence of Local Hidden Variable (LHV) is incompatible
with the outcome of experiments. A recent experiment \cite{r5} even showed
that the quantum correlation in an entangled state of two-photon system can
be maintained over long distance exceeding 10 km. However, it seems to us
that the original version of EPR paradox still remains to be answered.

The question EPR raised is as follows. As is well known in QM, the position
of a particle, $x$, and its momentum (in one dimensional space) operator

\begin{equation}
\hat{p}=-i\hbar \frac \partial {\partial x}  \label{e1}
\end{equation}
have a commutation relation: 
\begin{equation}
[x,\hat{p}]=i\hbar  \label{e2}
\end{equation}

Consider a two-particle system. Then the operators $(x_1-x_2)$ and $(\hat{p}
_1+\hat{p}_2)$ will commute: 
\begin{equation}
[x_1-x_2, \hat{p}_1+\hat{p}_2]=0  \label{e3}
\end{equation}
which means that they have a common eigenstate with eigenvalues 
\begin{equation}
x_1-x_2=a=const  \label{e4}
\end{equation}
and 
\begin{equation}
p_1+p_2=0,\quad p_2=-p_1  \label{e5}
\end{equation}

Now question arises: What strange the state is!? Two particles are moving
in opposite momentum directions while keeping their distance unchanged.
Incredible! As stressed in Ref. \cite{r2}: ``No one can figure out how to
realize it''. ``So there is no surprise that many people regarded the debate
(between EPR and Bohr) as an idle talk leading only to empty result''.
However, as we believe in the basic principles of QM, any difficulty must
have its root. Is this implying something overlooked or unfinished in QM?

In this letter, we wish to propose an answer to this EPR paradox, which
essentially prescribes the necessity of introducing the antiparticle state
in QM and even dictates the form of its Wave Function (WF) unambiguously.

If the WF of particle 1 is written as usual:

\begin{equation}
\psi (x_1,t)=\exp \{\frac i\hbar (p_1x_1-E_1t)\}  \label{e6}
\end{equation}

with momentum $p_1>0$ and energy $E_1>0$, then the particle 2 must be an
antiparticle described by the following WF:

\begin{equation}
\psi _c(x_2,t)=\exp \{\frac i\hbar (p_2x_2-E_2t)\}=\exp \{-\frac i\hbar
(p_1x_2-E_1t)\}  \label{e7}
\end{equation}

with $p_2=-p_1<0$, $E_2=-E_1<0$. But as proposed by Schwinger (\cite{r6}, see
also \cite{r7}) and stressed by one of the present authors, Ni (\cite{r8},
\cite{r9}), we should view the WF of ``negative energy state'' of particle
directly as the WF of its antiparticle with the corresponding antiparticle
operators: 
\begin{equation}
\hat{p}_c=i\hbar \frac \partial {\partial x},\hspace{0.5in}\hat{E}_c=-i\hbar \frac
\partial {\partial t}  \label{e8}
\end{equation}

So the observed momentum and energy of antiparticle in state
(\ref{e7}) are
$p_1$ and $E_1$ respectively, precisely the same as that of the particle
state (\ref{e6}). Therefore, the EPR state $\Psi =\psi  \psi_c$ is composed
of a particle-antiparticle pair with parallel momentum ($p_1$) and invariant
distance $(x_1-x_2)$ between them. Now every thing is reasonable.

In most text books on QM, there is nearly no clear explanation for WF of
antiparticle. Some times (and often implicitly), it is said that the WF of
antiparticle with momentum $p_1$ is the same as that of particle,
i.e.,
Eq. (\ref{e6}), one has to distinguish a positron from electron by adding some
words. Alternatively, one often said that in the vacuum all the negative
energy states of electron are filled. Once a ``hole'' is created in the
``sea'', it would correspond to a positron.

In accompanying with the historic discovery of parity (P) violation by
Lee-Yang \cite{r10} and Wu et. al. \cite{r11} in 1956, it was verified that the
charge conjugate transformation (C) which brings an electron (with charge
$-e$) to a positron (with charge $e$) is also violated \cite{r12}. In 1964, the
combined CP violation was discovered in $K^{0} -\bar{K^{0}}$ system
whereas the CPT theorem remains valid. The recent experiment by CPLEAR
Collaboration at CERN provides further direct observation of time-reversal
(T) non-invariance in the neutral-kaon system \cite{r13} (see also \cite{r14}
). What do all the above discoveries mean?

In our point of view, all the above discoveries imply that the individual
definition of P, C or T inversion ceases to be meaningful as an observable
transformation in physics. Of course, their definitions are still clearcut
in mathematics. Let us perform the combined CPT transformation on the WF of
a particle, either with or without spin. (see, e.g. in \cite{r15}). Because
the C or T transformation contains a complex conjugation respectively, they
cancel each other in the CPT transformation. Then what we obtain is essentially
 a transformation $(\vec{x}\rightarrow -\vec{x}, t\rightarrow -t)$
 while at the same time we should look at the WF describing an
 antiparticle. 
This is nothing but a statement of transforming Eq. (\ref{e6}) to
Eq. (\ref{e7}). 
And this is why physicists have been gradually accepting the relation
of a particle $|a \rangle$ 
with respect to its antiparticle $|\bar{a} \rangle$ being \cite{r16}

\begin{equation}
|\bar{a} \rangle=CPT|a \rangle  \label{e9}
\end{equation}

The reason why the observation Eqs. (\ref{e6})-(\ref{e8}) had not attracted
enough attention in physics community for so long a time is of two fold.
First, the inertia of concept in physics proves its strength once again as
witnessed in the history of science from time to time. Second, the
importance of the above observation has not fully explored until its
evolving into the following postulate: ``The space-time reversal $(\vec{x}
\rightarrow -\vec{x},t\rightarrow -t)$ transformation is equivalent to the
transformation between particle and antiparticle.'' This is a basic symmetry
which should be respected in constructing the theory for all particles and
fields. Thus a particle is always not pure. A particle state $\theta (\vec{x}
,t)$ is always accompanied by its antiparticle state $\chi (\vec{x},t)$ .
They have the simple relation:

\begin{equation}
\theta (-\vec{x},-t)=\chi (\vec{x},t)  \label{e10}
\end{equation}

The coupled equations should be invariant with respect to the transformation 
$(\vec{x}\rightarrow -\vec{x},t\rightarrow -t)$ together with Eq. (\ref{e10}).

Based on this symmetry, the essence of special relativity (SR) is explored
and the one-body relativistic equation, i.e., the Dirac equation and
Klein-Gordon equation, are derived \cite{r9}. Moreover, even to our
surprise, we began to realize that the Stationary Schr\"odinger Equation
(SSE) for many-body system is essentially relativistic as long as the
eigenvalue in SSE

\begin{equation}
H\psi =\varepsilon \psi  \label{e11}
\end{equation}
(say for two-body case, $H=p^2/2\mu +V(r)$ with $\mu$ being the reduced mass) is
related to the binding energy $B$ ($B\equiv M-E$ with $M$ and $E$ being the
rest mass and energy of whole system, $c=1$) as follows (\cite{r17}, \cite{r18}):

\begin{equation}
\varepsilon \equiv \frac{E^2-M^2}{2M},\quad B=M[1-(1+\frac{2\varepsilon }
M)^{1/2}]  \label{e12}
\end{equation}

It is interesting to see that the relativistic correction in Eq. (\ref{e12})
for Hydrogenlike atom is a small downward energy shift

\begin{equation}
\Delta E^{rel}=-\frac{\varepsilon ^2}{2M},\quad (\varepsilon =-\frac{
Z^2\alpha ^2\mu }{2n^2})
\end{equation}
which only accounts for ($-23.814MHz$) , or ($-0.3\%$) in the magnitude of
the so-called absolute Lamb shift (upward) $8172.86MHz$ of 1S state in
Hydrogen atom \cite{r19}. Based on the new understanding of SSE, Eq (\ref
{e11}), a simple noncovariant calculation on Lamb shift is performed in Ref
\cite{r20} with high accuracy ($<0.1\%$).

In summary, some discussions are in order:

(a) The original version of EPR paradox was a simple but acute problem
querying on whether the QM is complete. If instead of Eq. (\ref{e3}) we
consider

\begin{equation}
\lbrack x_1+x_2,\hat{p}_1-\hat{p}_2]=0  \label{e14}
\end{equation}

Then the correct answer turns out to be a particle and its
antiparticle ($c$)
moving in opposite direction with momentum $p_1$ and $p_2^{(c)}=-p_1$ and
position $x_1$ and $x_2=-x_1$. Such kind of experiments have been performed
for many times. For example, the $e^{+}$ and $e^{-}$ pair can be created by
a high energy photon in the vicinity of heavy nuclei. The $0^{+}$ excited
state of $^{16}O$ nuclei lies higher $6MeV$ than the ground $0^{+}$ state,
so it can also give rise to pair creation of $e^{+}$ and $e^{-}$. Recently,
the experiment at CERN again demonstrated the quantum correlation at a
distance for kaon and antikaon system. They are entwined. Only until the
measurement on $p_1$ direction to see a particle (or antiparticle), can one
predict with 100 percent probability that an antiparticle (or particle)
appearing in the $p_2$ direction \cite{r21}.

All EPR experiments in the past 64 years have been proving the validity and
completeness of QM, excluding the existence of LHV. What we add in this
letter is that some times the including of antiparticle is necessary.
Actually, the correct answer to the original version of EPR paradox is a
unique solution. There is no other alternative.

(b) The cognition of Eqs. (\ref{e6})-(\ref{e9}) is in total conformity with
the long investigation on the problem of C, P and T reversal since the
discovery of parity violation in 1956. In other words, the CPT theorem
already exhibits itself as a basic postulate. The transformation of a
particle to its antiparticle is not some thing which can be defined
independently but a direct consequence of (newly defined) space-time
reversal $(\vec{x}\rightarrow -\vec{x},t\rightarrow -t)$ .

(c) The above basic symmetry should be pushed forward into Eq. (\ref{e10}),
forming a starting point to construct the SR, the relativistic QM and the
quantum field theory (QFT). (\cite{r8}, \cite{r9} see also \cite{r22}).

(d) What we are discussing is the essential conformity between QM and SR. In
some sense the subtle relationship
between them lies in the essential equivalence of ``$%
+i$'' and ``$-i$''. Indeed, at the very elementary level of physics, ``that
not forbidden is allowed'' (M. Gell-Mann).

\section*{Acknowledgments} 
We thank Profs. M-l Du, H-z Li, J-q Liang, R-k Su , B-w
Xu, S-q Ying, X-y Zeng and Drs. G-h Yang, J-f Yang and Z Zhang for
discussions. This work was supported in part by the NSF of China.

\end{document}